# Thermodynamic and corrosion study of $Sm_{1-x}Mg_xNi_y$ ($y$ = 3.5 or 3.8) compounds forming reversible hydrides

V. Charbonnier, N. Madern, J. Monnier, J. Zhang, V. Paul-Boncour, M. Latroche

Univ. Paris Est Creteil, CNRS, ICMPE, UMR7182, F-94320, Thiais, France.

**Abstract**

$AB_5$ compounds ($A$ = rare earth, $B$ = transition metal) have been widely studied as anodes for Ni-$M$H applications. However, they have reached their technical limitations and the search for new promising materials with high capacity is foreseen. $AB_y$ compounds (2 < $y$ < 5) are good candidates. They are made by stacking [$AB_5$] and [$A_2B_4$] units along the $c$ crystallographic axis. The latter unit allows a large increase in capacity, while the [$AB_5$] unit provides good cycling stability. Consequently, the $AB_{3.8}$ composition (*i.e.* $A_5B_{19}$ with three [$AB_5$] for one [$A_2B_4$]) is expected to exhibit better cycling stability than the $AB_{3.5}$ (*i.e.* $A_2B_7$ with two [$AB_5$] for one [$A_2B_4$]). Furthermore, substitution of rare earth by light magnesium improves both the capacity and cycling stability. In this paper, we compare the hydrogenation and corrosion properties of two binary compounds $SmNi_{3.5}$ and $SmNi_{3.8}$ and two pseudo-binary ones (Sm,Mg)$Ni_{3.5}$ and (Sm,Mg)$Ni_{3.8}$. A better solid-gas cycling stability is highlighted for the binary $SmNi_{3.8}$. The pseudo-binary compounds also exhibit higher cycling stability than the binary ones. Furthermore, their resistance to corrosion was investigated.

I. **Introduction**

The capacity of $AB_5$ compounds ($A$ = rare earth, $B$ = transition meta), used as anode material in Ni-$M$H batteries reaches 300 to 350 mAh/g [1, 2, 3, 4] [6, 7]. However, they are now approaching their upper limit. Due to the overgrowing energy need, new electrode materials are deeply investigated worldwide in the view of increasing weight capacity of Ni-$M$H batteries. Active material for negative electrode must have specific properties: ability to absorb reversibly a large amount of hydrogen, excellent resistance to corrosion, high kinetics of absorption and desorption of hydrogen and high cycling stability [5]. Consequently, the *P-c* isotherms of these materials must have a large and flat plateau pressure, lying in the practical electrochemical window (0.001 to 0.1 MPa at room temperature) [1].

In the past twenty years, the $AB_y$ compounds with stacking structures (2 < $y$ < 5) [8, 9, 3, 10] have attracted lot of interests due to their high capacity. They can be considered by piling [$A_2B_4$] and [$AB_5$] units along the $c$ crystallographic axis. $AB_y$, according to the following formula introduced by Khan in 1974 [11], can be described as stacking of $n$[$AB_5$] and one [$A_2B_4$], were $y = \frac{5n+4}{n+2}$. For $AB_3$, $n$ equals 1 and the basic stacking period is 1x[$AB_5$]+[$A_2B_4$]; for $A_2B_7$, $n$ equals 2 and the basic stacking period is 2x[$AB_5$]+[$A_2B_4$], whereas for $A_5B_{19}$, $n$ equals 3. $AB_y$ phases are polymorphic as they can adopt either rhombohedral or hexagonal symmetry. For rhombohedral (hexagonal) structure, the basic stacking period is repeated three (two) times along the $c$-axis respectively. The [$A_2B_4$] unit brings higher capacity when the [$AB_5$] unit provides better stability upon hydrogenation/dehydrogenation cycling [1]. Thus, $AB_3$ compounds are known to exhibit good capacity as demonstrated in [12] but rather poor cycle life [13, 14, 15]. In the present work, we focus on $A_2B_7$ and $A_5B_{19}$ compounds because the higher $n$ ratio provides the best tradeoff between stability and capacity . Interestingly, part of the rare earth *A* can be substituted by magnesium within the [$A_2B_4$] unit. Its low molar mass and strong affinity for hydrogen lead to a large increase of the weight specific capacity (around 400 mAh/g [9]).

For the *A* side, lightest rare earths (La, Pr and Nd) are usually chosen but they are also the largest ones according to the lanthanide contraction. It has been shown that for $AB_2$ compounds with Laves phase structures, the more stable atomic radius ratio $R_A/R_B$ should be close to 1.225. For rare earths with large radii, this ratio is much higher leading to strong lattice strains inducing amorphization upon hydrogenation [16, 17, 18, 19, 20]. Aoki *et al.* demonstrated experimentally that this ratio should be below 1.37 [21, 16] to avoid this phenomenon whereas the $R_A/R_B$ ratio ranges between 1.50 and 1.46 for the early mentioned rare earths. To decrease this ratio and to broaden the knowledge on $AB_y$ compounds, we have investigated the $SmNi_y$ system ($R_A/R_B$ = 1.45) with *y* = 3.5 (*n* = 2) and *y* = 3.8 (*n* = 3). Then, the Mg substitution for Sm is studied. Thermodynamic and corrosion properties are described at the light of the composition changes.

## II.   Experiments

$Sm_2Ni_7$, $Sm_{1.65}Mg_{0.35}Ni_7$, $Sm_5Ni_{19}$ and $Sm_{4.5}Mg_{0.5}Ni_{19}$ have been synthesized from high purity elements (Sm (Alfa Aesar 99.9%), Mg (Alfa Aesar 99.8%), Ni (Praxair 99.95%)). The binary compounds were prepared with a small excess of samarium to account for Sm sublimation upon heating. After induction melting under argon, the ingots were crushed into powder (< 100 µm). They were pressed into a 2 g-pellet, wrapped in tantalum foil and annealed under argon in a sealed stainless steel crucible for 3 days at 950 °C. For the Mg-containing compounds, a Sm-Ni precursor was first prepared by induction melting. The obtained ingot was crushed into powder (< 100 µm). Then magnesium powder (< 44 µm) was added in small excess. The final powder was pressed into 2 g-pellets, which were annealed following the same procedure as for the binary compounds.

Chemical composition was checked by Electron Probe Micro-Analysis (EPMA) with a CAMECA SX-100. The samples were mechanically polished to 0.25 µm using diamond paste. Structural properties were characterized by X-ray diffraction (XRD) using a Bruker D8 DAVINCI diffractometer with Cu-Kα radiation, in a 2$\theta$-range from 20 to 80° with a step size of 0.01°. Experimental data were analyzed by the Rietveld method using the FullProf program [22].

Pressure-composition (*P-c*) isotherms were measured at 25 °C using Sieverts' method followed by desorption under dynamic vacuum [23]. The compounds were crushed into 100 µm-powder under argon atmosphere and 500 mg of sample were used for the measurements. For the samples containing magnesium, a *P-c* isotherm with maximal pressure of 10 MPa followed by five activation cycles were performed. An activation cycle consists in absorption at 2.5 MPa and desorption under dynamic vacuum at 150 °C.

For solid-gas cycling, each cycle starts by introducing hydrogen in the sample holder initially under vacuum. When equilibrium is reached, the amount of absorbed hydrogen is calculated. The initial hydrogen pressure was determined so that the equilibrium pressure reaches 0.1 MPa. Between each cycle, desorption was performed at 150 °C under dynamic vacuum for several hours.

For corrosion studies, several sample holders were used. In each of them, 200 mg of powdered sample (< 100 µm) were immersed in same quantity of a 8.7 M KOH solution (electrolyte). The sample holders were kept in an oven with regulated temperature of 25 °C under continuously renewed argon atmosphere. After a given time *t*, the electrolyte was removed with a Pasteur pipette and the powder was rinsed several times with a $10^{-2}$ M KOH solution. The rinsing solution has a *p*H of 12 to allow removing of the potassium hydroxide without dissolving the corrosion products. Then, the powder was dried at 40 °C under dynamic vacuum for 24 h and analyzed by XRD and magnetic measurements.

XRD analysis is useful to determine the nature of the corrosion products. However, the size of the initial powder is very coarse (< 100 μm), so that the diffraction of X-rays occurs mainly at the surface of the sample [24]. Consequently, XRD analysis is not well adapted to evaluate the phase percentage of the corroded samples. Among the expected corrosion products only metallic Ni is ferromagnetic at room temperature, whereas the others are paramagnetic. Therefore, the amount of metallic nickel in the corroded samples can be estimated by magnetic measurements [25, 26].

Isotherm magnetization curves were measured at 300 K with a Physical Properties Measurement System (PPMS) from Quantum Design. Applied magnetic field was varied from 0 to 9 T. A small amount of sample (20 to 50 mg) was fixed by a resin in a Teflon sample holder, itself introduced in a gelatin capsule and fixed with glass wool. The diamagnetic contribution of the sample holder was measured and substracted from the sample magnetization. Assuming that the ferromagnetic contribution comes exclusively from Ni, the extrapolation of the magnetization curves at zero field and the saturation magnetization of nickel were used to calculate the weight percent of Ni in the corroded samples.
The mass percentage of Ni is then obtained by normalization with the saturation magnetization of nanoporous nickel isolated from a fully corroded sample (42.74 $Am^2\ kg^{-1}$ at 300 K compared to 53 $Am^2\ kg^{-1}$ for bulk Ni).

III. Results

1- **Phase determination and structural properties**

EPMA was performed to check the chemical composition of $Sm_2Ni_7$, $Sm_{1.65}Mg_{0.35}Ni_7$, $Sm_5Ni_{19}$ and $Sm_{4.5}Mg_{0.5}Ni_{19}$. The main phase of the four studied compounds corresponds to the targeted composition (Figure S1). Figure 1 (a) shows the XRD patterns of $Sm_2Ni_7$, $Sm_{1.65}Mg_{0.35}Ni_7$, $Sm_5Ni_{19}$ and $Sm_{4.5}Mg_{0.5}Ni_{19}$ compounds. Rietveld analysis for $Sm_{4.5}Mg_{0.5}Ni_{19}$ is also presented (Figure 1 (b)), the Rietveld analysis for the other three compounds are presented in Figure S2. $Sm_2Ni_7$ and $Sm_{1.65}Mg_{0.35}Ni_7$ are purely $A_2B_7$-type alloys with $Gd_2Co_7$-type and $Ce_2Ni_7$-type polymorphs. For $Sm_5Ni_{19}$ and $Sm_{4.5}Mg_{0.5}Ni_{19}$ compounds, the main phases are $A_5B_{19}$-type with polymorphic forms: rhombohedral $Ce_5Co_{19}$-type structure and hexagonal $Sm_5Co_{19}$-type structure. Small quantities of $SmNi_5$ are also present. Weight percentage of the observed phases and cell parameters derived from Rietveld refinement are summarized in Table 1. The substituted compounds have smaller cell parameters as compared to the binary ones due to the smaller radius of Mg. Mg substitutes Sm in the [$A_2B_4$] unit only. The occupation rate *x* of Mg is given in Table 1. The average compositions derived from Rietveld refinement are calculated as $Sm_{1.65(2)}Mg_{0.34(3)}Ni_7$ and $Sm_{4.56(1)}Mg_{0.43(2)}Ni_{19}$, in good agreement with EPMA analysis (Figure S1).

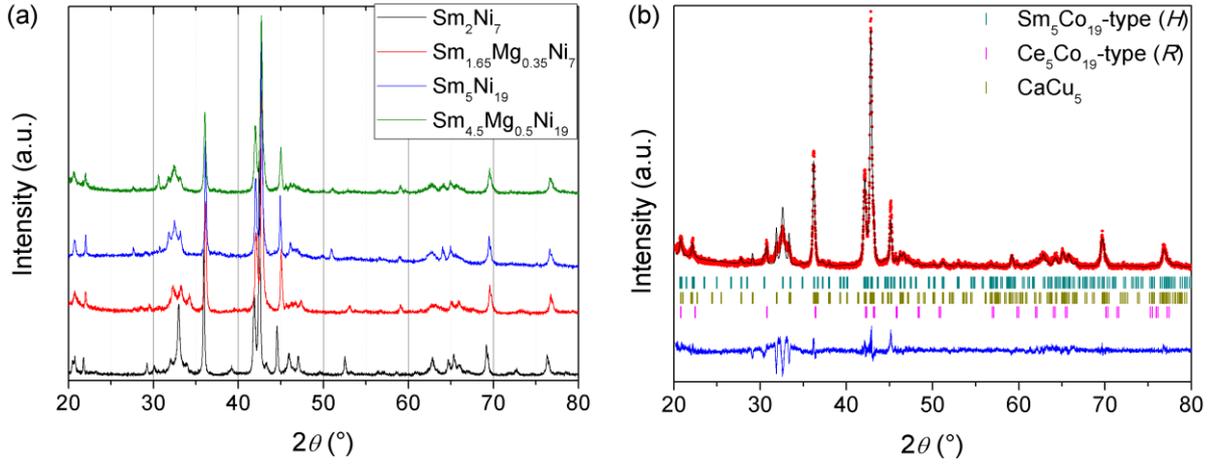

Figure 1: XRD patterns for $Sm_2Ni_7$, $Sm_{1.65}Mg_{0.35}Ni_7$, $Sm_5Ni_{19}$ and $Sm_{4.5}Mg_{0.5}Ni_{19}$ (a) and Rietveld analysis of the XRD pattern for $Sm_{4.5}Mg_{0.5}Ni_{19}$ (b).

Table 1: Crystallographic parameters of studied compounds derived from Rietveld refinement and their evolution after hydrogenation cycling under 10 MPa. *R*: rhombohedral structure, *H*: hexagonal structure, *x*: refined occupation rate of Mg (exclusively located in [$A_2B_4$] unit)

|  | Pristine compounds | | | Evolution of the cell parameters of the cycled compounds | |
|---|---|---|---|---|---|
|  | *R* $R\bar{3}m$ | *H* $P6_3/mmc$ | $CaCu_5$-type $P6/mmm$ | *R* $R\bar{3}m$ | *H* $P6_3/mmc$ |
| **$Sm_2Ni_7$** | 21 wt.% | 79 wt.% | 0 wt.% |  |  |
|  | a=4.976(2) Å | a=4.9809(5) Å |  | Δa/a=-0.12(1)% | Δa/a=0.04(5)% |
|  | c=36.55(1) Å | c=24.322(4) Å |  | Δc/c=0.40(1)% | Δc/c=1.66(6)% |
| **$Sm_{2-x}Mg_xNi_7$ (x = 0.35)** | 71 wt.% | 29 wt.% | 0 wt.% |  |  |
|  | a=4.9607(5) Å | a=4.9625(7) Å |  | Δa/a=0.35(3)% | Δa/a=0.50(3)% |
|  | c=36.167(4) Å | c=24.117(4) Å |  | Δc/c=1.32(3)% | Δc/c=1.16(4)% |
|  | x = 0.39(2) | x = 0.21(3) |  |  |  |
| **$Sm_5Ni_{19}$** | 51% | 47% | 2 wt.% |  |  |
|  | a=4.9698(3) Å | a=4.9694(3) Å | a= 4.937(3) Å | Δa/a=-0.06(2)% | Δa/a=-0.16(2)% |
|  | c=48.401(3) Å | c=32.258(2) Å | c= 3.964(4) Å | Δc/c=0.82(3)% | Δc/c=0.38(4)% |
| **$Sm_{5-x}Mg_xNi_{19}$ (x = 0.5)** | 64 wt.% | 30 wt.% | 6 wt.% |  |  |
|  | a=4.9610(5) Å | a=4.9556(4) Å | a= 4.939(1) Å | Δa/a=0.10(2)% | Δa/a=0.12(2)% |
|  | c=48.190(5) Å | c=32.089(5) Å | c= 3.962(1) Å | Δc/c=0.54(3)% | Δc/c=0.60(4)% |
|  | x = 0.49(1) | x = 0.40(2) |  |  |  |

2- **Hydrogenation properties**

*P-c* isotherms. *P-c* isotherms of $Sm_2Ni_7$, $Sm_{1.65}Mg_{0.35}Ni_7$, $Sm_5Ni_{19}$ and $Sm_{4.5}Mg_{0.5}Ni_{19}$ were measured using Sieverts' method (Figure 2 (a) and (b)). The four compounds exhibit two pressure plateaus. For $Sm_2Ni_7$, the first one appears at 0.01 MPa and is lying between 0.01 and 0.41 H/*M* (H/*M*: hydrogen per metal atom), the second one is approximately at 7 MPa and starts at 0.57 H/*M*. For $Sm_{1.65}Mg_{0.35}Ni_7$, the first plateau is a sloping one near 0.4 MPa ranging between 0.06 to 0.56 H/*M*, the second one is around 10 MPa and starts at 0.78 H/*M*. The end of this plateau exceeding 10 MPa was not measured. The first plateau of $Sm_5Ni_{19}$ starts at 0.02MPa for a capacity range of 0.02 to 0.33 H/*M*. The second one is located around 5.6 MPa for a capacity between 0.44 and 1.17 H/*M*. For $Sm_{4.5}Mg_{0.5}Ni_{19}$, the first plateau appears at 0.9 MPa, it corresponds to a capacity between 0.04 and 0.98 H/*M*.; the second one starts around 10 MPa, for a capacity of 1.06 H/*M*. Furthermore, Figure 2 shows that $Sm_{1.65}Mg_{0.35}Ni_7$, $Sm_5Ni_{19}$ and $Sm_{4.5}Mg_{0.5}Ni_{19}$ desorb hydrogen, whereas $Sm_2Ni_7$ does not.

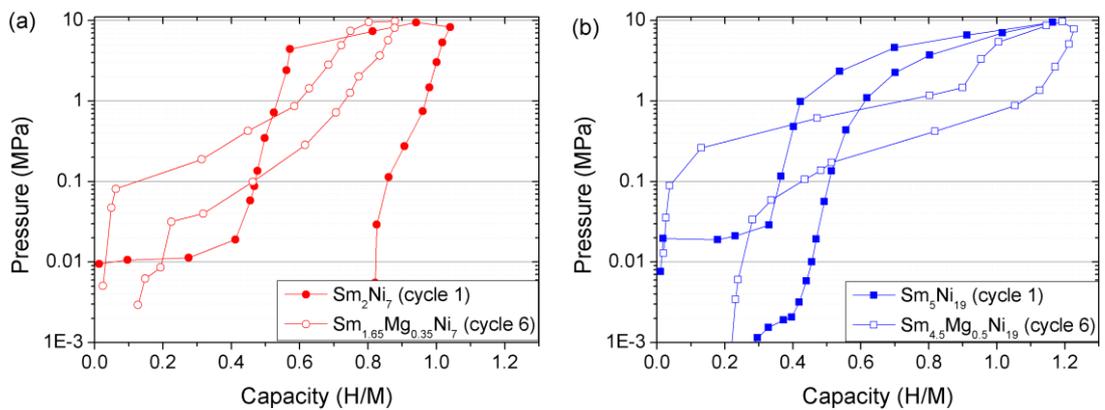

**Figure 2:** Comparison of *P-c* isotherms measured at 25°C for $Sm_2Ni_7$ and $Sm_{1.65}Mg_{0.35}Ni_7$ (a) and $Sm_5Ni_{19}$ and $Sm_{4.5}Mg_{0.5}Ni_{19}$ (b).

**Structure stability upon hydrogenation.** XRD patterns measured before and after *P-c* isotherm measurements of $Sm_2Ni_7$, $Sm_{1.65}Mg_{0.35}Ni_7$, $Sm_5Ni_{19}$ and $Sm_{4.5}Mg_{0.5}Ni_{19}$ are presented in Figure 3. For the four compounds, the structure is preserved after dehydrogenation, but peak broadening is observed (especially for $Sm_2Ni_7$) and diffraction peaks are shifted. The cell parameters of the cycled compounds were determined by Rietveld refinement. The differences with the pristine compounds are given in Table 1.

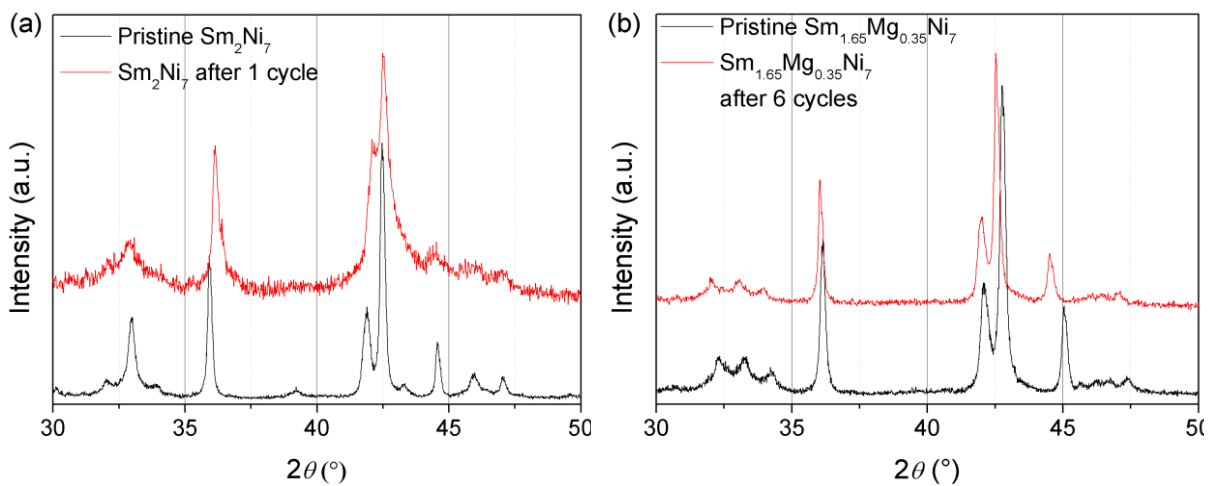

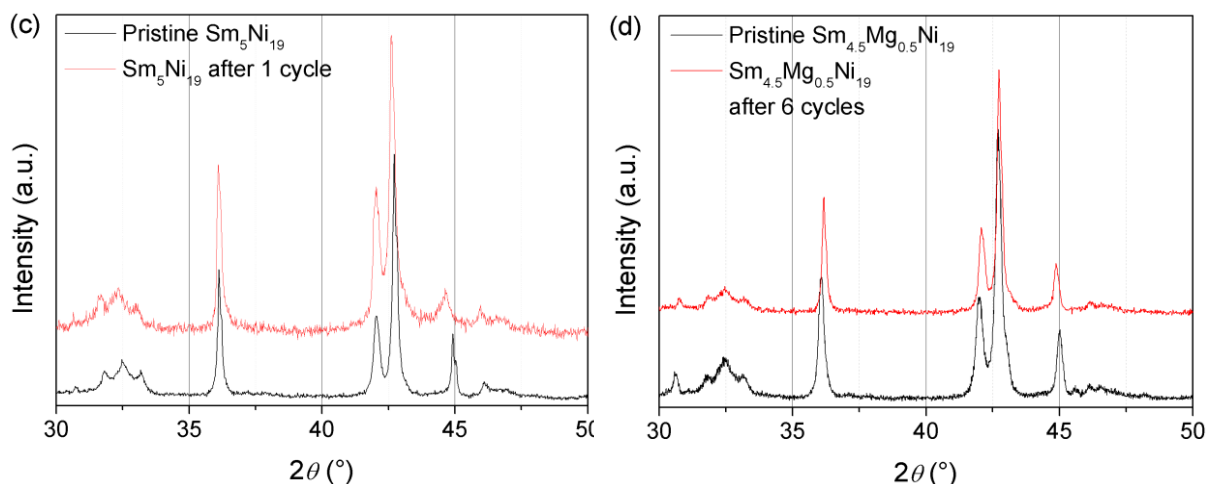

Figure 3: XRD patterns of pristine and dehydrogenated compounds under dynamic vacuum at 150 °C: $Sm_2Ni_7$ (a), $Sm_{1.65}Mg_{0.35}Ni_7$ (b), $Sm_5Ni_{19}$ (c) and $Sm_{4.5}Mg_{0.5}Ni_{19}$ (d). The maximal pressure applied was 10 MPa at 25 °C.

Solid-gas cycling corresponding to the first pressure plateau was performed for $Sm_2Ni_7$ and $Sm_5Ni_{19}$ (the maximal pressure applied was calculated so that the theoretical equilibrium pressure was 0.1 MPa). Figure 4 (a) shows the evolution of the capacity upon cycling. The capacity drastically decreases for $Sm_2Ni_7$ after 10 cycles, whereas only a slight diminution is observed for $Sm_5Ni_{19}$ after 20 cycles. XRD patterns measured after solid-gas cycling show that $Sm_2Ni_7$ presents large peak broadening, whereas the diffraction peaks remained narrow for $Sm_5Ni_{19}$ (Figure 4 (b))

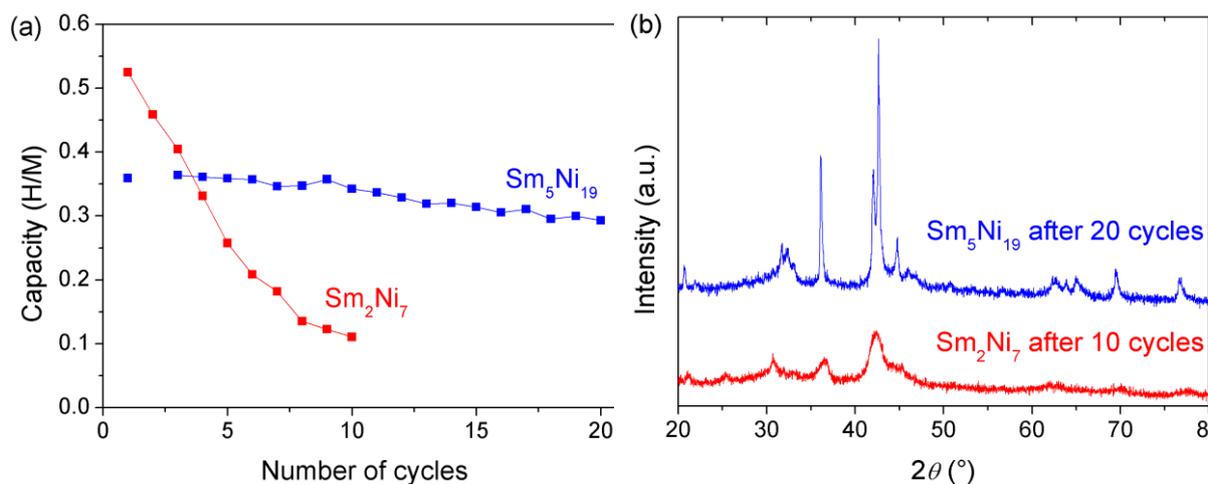

Figure 4: Solid-gas cycling of $Sm_2Ni_7$ and $Sm_5Ni_{19}$ performed at 25°C with a theoretical equilibrium pressure of $H_2$ of 0.1 MPa (a) and XRD patterns recorded after solid-gas cycling (b).

3- Corrosion mechanism

**X-ray diffraction analysis.** Figure 5 compares the XRD patterns of the binary and pseudo-binary compounds after different corrosion times in KOH solution at room temperature. For each compound, the progressive appearance of diffraction peaks of samarium hydroxide $Sm(OH)_3$ with *P6$_3$/m* space group and cubic nickel with *Fm-3m* space group is observed. These corrosion products are coherent with previous studies [25, 26]. Thorough XRD analysis did not allow the detection of $Ni(OH)_2$ or $Mg(OH)_2$.

For the compounds corroded 8 weeks, a quick comparison of the relative intensity of diffraction peaks of the pristine structure and Sm(OH)$_3$ shows that the peuso-binaries exhibit less corrosion products compared to the binaries. After 18 weeks of corrosion, no clear difference is observed.

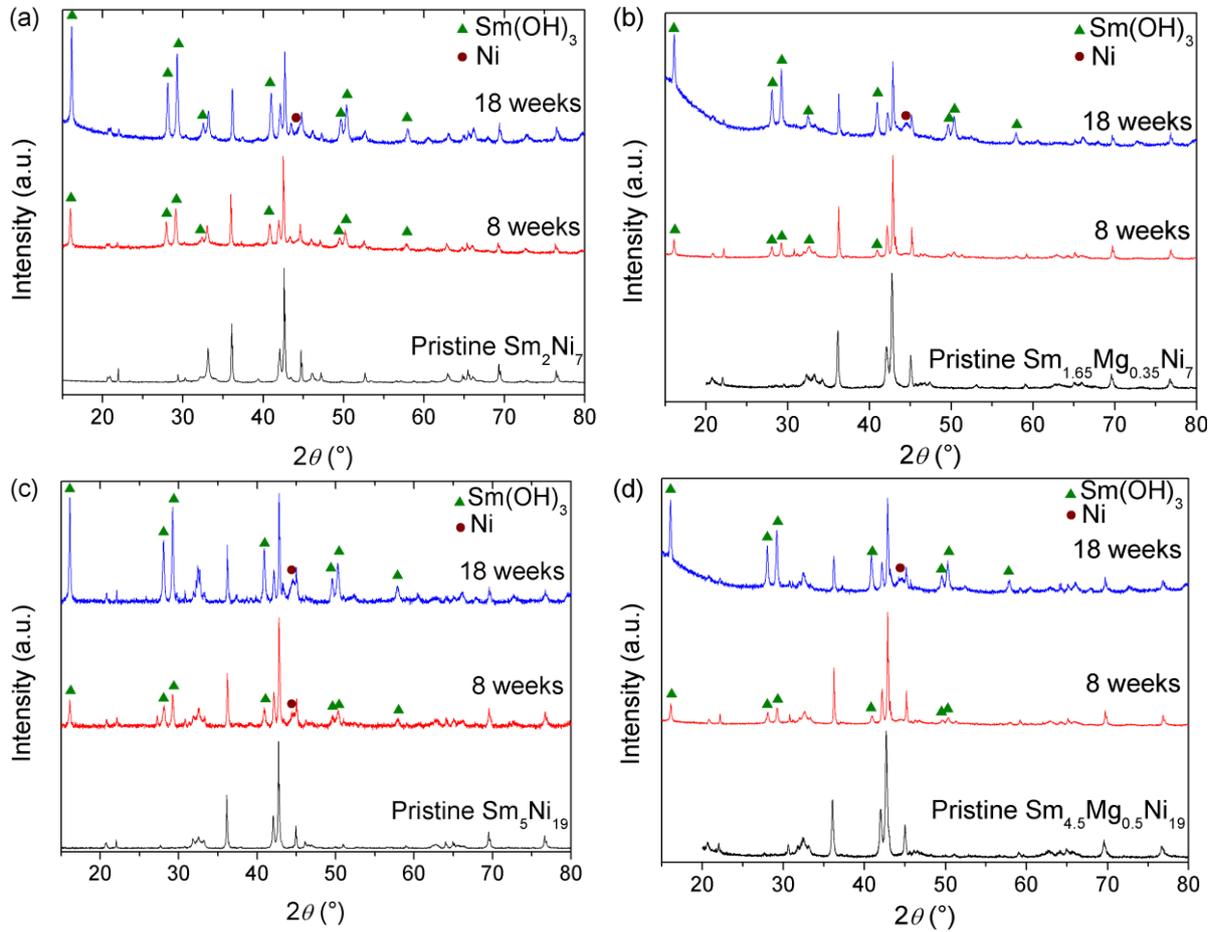

Figure 5: XRD patterns of the corroded samples: Sm$_2$Ni$_7$ (a), Sm$_{1.65}$Mg$_{0.35}$Ni$_7$ (b), Sm$_5$Ni$_{19}$ (c) and Sm$_{4.5}$Mg$_{0.5}$Ni$_{19}$ (d).

**Magnetic analysis.** The evolution of the weight percent of metallic Ni obtained by magnetic measurement is summarized in Figure 6. It shows that the pristine alloys do not contain ferromagnetic Ni metal. However, Ni appears and its quantity increases with time. Weight percent of Sm$_{1-x}$Mg$_x$Ni$_y$ lost during corrosion was estimated from the abundance of Ni metal assuming that the reaction is the following:

$$2\,Sm_{1-x}Mg_xNi_y + 2(3-x)\,H_2O$$
$$\rightarrow 2(1-x)Sm(OH)_3 + 2x\,Mg(OH)_2 + 2y\,Ni + (3-x)\,H_2 \qquad (1)$$

The results are plotted in Figure 6. For short corrosion times, Sm$_5$Ni$_{19}$ corrodes slowly compared to Sm$_2$Ni$_7$. However, after 8 weeks the amount of corroded alloy is similar for both binary compounds and saturates around 20 wt%. For the pseudo-binary compounds, the 5:19 corrode slowly compared to the 2:7. Although the amount of corroded compound is lower after 8 weeks compared to binary ones, it increases a lot and after 18 weeks it reaches the same level for 5:19 compounds and becomes higher for the binary one for 2:7 compound.

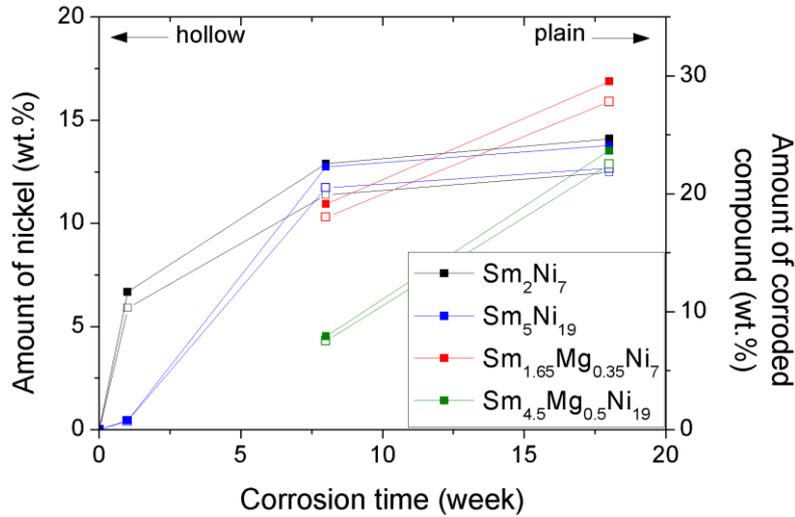

Figure 6: Weight percent of nickel derived from magnetic measurements performed after different corrosion times in KOH solution (8.7M) for $Sm_2Ni_7$, $Sm_{1.65}Mg_{0.35}Ni_7$, $Sm_5Ni_{19}$ and $Sm_{4.5}Mg_{0.5}Ni_{19}$.

## IV. Discussion

### 1. Corrosion

The corrosion products identified in this study are the same for the binary and pseudo-binary compounds, *i.e.* $Sm(OH)_3$ and metallic Ni. These products are in good agreement with previous observations [25, 26, 27]. Other corrosion products that could possibly be formed in the present conditions are: nickel hydroxide and/or oxide [27, 28] and magnesium hydroxide [28]. None of them was identified even after thorough XRD analysis. Indeed, their amount should be very low, since Mg content is small (between 2 to 4 at.%) and even after 18 weeks, only a small amount of alloy is corroded. The amount of corrosion products for Mg is accordingly low, making it difficult to detect. More local techniques such as Raman spectroscopy or transmission electron microscopy might be used for further investigation. Since no nickel hydroxide and/or oxide was identified, we assumed in Equation (1) that the corrosion product is only metallic Ni.

Compared to binary compounds, corrosion of Mg-substituted compounds is slow at short times. Then the corrosion rate increases and becomes faster than for the binary compounds. Cai *et al.* studied the effect of Mg in $La_{1-x}Mg_xNi_{1.75}Co_{2.05}$ and concluded that increasing the Mg content accelerates the corrosion of the alloy [29].

It is also worth to compare the corrosion behavior of 2:7 and 5:19 systems. For short times, the corrosion is more important for $Sm_2Ni_7$ than for $Sm_5Ni_{19}$, but after 8 weeks, the amount of metallic Ni is the same for both compounds (Figure 6). Sm is not stable in basic medium, especially compared to Ni, and the Sm content in $Sm_2Ni_7$ (22.22 at.%) is larger than in $Sm_5Ni_{19}$ (20.83 at.%). Hence the corrosion is faster for $Sm_2Ni_7$ at short times. For long time (after 8 weeks here), the corrosion could be inhibited by a passivation layer. Similar results are observed for the Mg-substituted. It was mentioned earlier that Mg enhances corrosion and $Sm_{1.65}Mg_{0.35}Ni_7$ presents a slightly higher Mg content (4 at.%) compared to $Sm_{4.5}Mg_{0.5}Ni_{19}$ (2 at.%). The higher Mg and *A*-element contents in $A_2B_7$ explain the significantly larger corrosion of $Sm_{1.65}Mg_{0.35}Ni_7$ compared to $Sm_{4.5}Mg_{0.5}Ni_{19}$.

## 2. Effect of the substitution of Sm by Mg

**Structural properties.** $AB_y$ (3<*y*<5) are polymorph structures that crystallize in rhombohedral (*R*) with *R3̄m* or hexagonal (*H*) with *P6$_3$/mmc* space group [30, 11].

Crivello *et al.* determined the preferential site occupation of Mg in $La_{1-x}Mg_xNi_y$ (*y* = 3.5 and 3.8) by DFT calculations [30]. They showed that Mg preferentially occupies the [$A_2B_4$] unit. They attributed this behavior to the lower radius of Mg, which prefer to occupy low coordination sites (*A* site in [$A_2B_4$], coordination number CN=16) whereas larger La occupies higher coordination sites (*A* site in [$AB_5$], CN= 20). In the same line, our Rietveld refinements show that Mg substitutes Sm in the [$A_2B_4$] unit only, in good agreement with other experimental studies [33, 34, 35].

**Equilibrium plateau pressure..** The comparison of the P-c isotherms for $Sm_2Ni_7$, $Sm_{1.65}Mg_{0.35}Ni_7$, $Sm_{1.6}Mg_{0.4}Ni_7$ [36] and $Sm_{1.5}Mg_{0.5}Ni_7$ [37] shows that an increase in the substitution of Sm by Mg leads to an increase of the first plateau pressure (Figure S3). As Mg is smaller than Sm, the cell volume for the Mg-substituted compounds is smaller than for the binary ones ($Sm_2Ni_7$ or $Sm_5Ni_{19}$), leading to a reduction of the size of the interstitial sites for hydrogen. As pointed out by Mendelsohn *et al.* [38], ln *P* is directly proportional to the inverse of the cell volume (with *P* the plateau pressure). Consequently a smaller cell volume generates a higher plateau pressure. Furthermore, the plateau pressure is higher for the formation of $MgH_2$ ($\Delta H$ = -74.5 kJ/mol of $H_2$ [39]) than for the formation of $SmH_2$ ($\Delta H$ = -202.6 kJ/mol of $H_2$ [40]). Compared to Sm, the small radius of Mg as well as the high enthalpy of hydride formation both explain the increase of the first plateau pressure.

Beside equilibrium pressure, it is worth to note that Mg generates larger and flatter plateau. Denys *et al.* [33] studied the evolution of the crystallographic parameters upon deuteration of $La_2Ni_7$ and $La_{1.5}Mg_{0.5}Ni_7$. They noticed that cell expansion induced by absorption of deuterium is anisotropic along the *c*-axis for the binary compound and isotropic for the Mg-substituted one. Anisotropic expansion was also observed upon hydrogenation of other $ANi_y$ binary compounds (*y* = 3 or 3.5) after absorption of a small amount of D [41, 42, 43, 44, 45]. Neutron diffraction have shown that, at low D content, the deuterium only enters [$A_2B_4$] units. [$AB_5$] units prevent expansion in the basal (*ab*)-plan leading to anisotropic expansion along the *c*-axis. Similarly, isotropic expansion, observed for Mg-substituted compounds [33], suggests that [$A_2B_4$] and [$AB_5$] units are filled together involving a larger number of insertion sites for H(D). This may explain the larger pressure plateau observed for the Mg-substituted compounds compared to the binary ones (Figure 2).

In the same line, Gal *et al.* compared the hydrogenation properties of $La_2Ni_7$ and Mg-substituted $La_{1.5}Mg_{0.5}Ni_7$, at 25 °C [46]. The binary compound exhibits two pressure plateaus. The first one lays at very low pressure with a maximal hydrogen content of 0.45 H/*M*. In contrast, the substituted compound exhibits a single plateau pressure at 0.024 MPa in the range 0.21 to 0.84 H/*M*. Compared to the first pressure plateau of $La_2Ni_7$, this plateau is larger and at higher pressure. This reinforces our hypothesis that for $AB_y$ compounds, the substitution of the rare earth by Mg leads to (i) an increase of the first plateau pressure (due to its small radius and its lower enthalpy of hydride formation compared to $A$) and (ii) an increase of the width of the first pressure plateau (due to the simultaneous filling of [$A_2B_4$] and [$AB_5$] units). This last point is particularly interesting since, as mentioned in the introduction, it is necessary to have large and flat pressure plateaus in the view of application as anode for Ni-$M$H batteries.

For the Sm-Mg-Ni system, the investigated compounds present an anisotropic expansion along the $c$-axis (Table 1) after $P$-$c$ isotherm measurement. Despite desorption under dynamic vacuum, a little amount of hydrogen remains in the in the solid solution phase, probably located in the [$A_2B_4$] units. This lattice expansion and the fact that hydrogen was not fully desorbed suggest that, similarly to other non Mg-substituted $AB_y$ compounds [42, 45] and though a longer pressure plateau is observed for Mg-substituted $AB_y$ (Figure 2), hydrogen insertion happens in two steps: first filling the [$A_2B_4$] units, then filling the whole structure. This is in line with the results of Zhang *et al.* [36], who investigated the structure evolution of $Sm_{1.6}Mg_{0.4}Ni_7$ upon hydrogenation and concluded that the [$A_2B_4$] unit was filled prior to the [$AB_5$] one. For the La-Mg-Ni system, the optimum Mg content to get a flat plateau is $La_{2-x}Mg_xNi_7$ with $x$ = 0.5 [47]. When increasing the Mg content, the mismatch between [$AB_5$] and [$A_2B_4$] units decreases. Upon hydrogenation of $Sm_{1.5}Mg_{0.5}Ni_7$, a single plateau is observed [37] (Figure S3). It would be interesting to investigate hydrogenation process by *in-situ* X-ray diffraction analysis and determine if the structure undergoes isotropic expansion (like $La_{1.5}Mg_{0.5}Ni_7$) or anisotrpic expansion.

**Cycling stability.** All the compounds preserved their structrure after $P$-$c$ isotherm measurements, but $Sm_2Ni_7$ undergoes a severe peak broadening, highlighting loss of crystallinity of the structrure (Figure 3 (a)). Furthermore, desorption of hydrogen was not possible under $P$-$c$ isotherm condition unlike $Sm_{1.65}Mg_{0.35}Ni_7$ (Figure 2 (a)). For the 5:19, $Sm_5Ni_{19}$ presents larger peak broadening than $Sm_{4.5}Mg_{0.5}Ni_{19}$ after dehydrogenation (Figure 3 (c) and (d)). It indicates that the structural stability upon cycling is better for the Mg-substituted $AB_y$. This is in line with the results of Denys *et al.*, who compared the hydrogenation properties of the binary $La_2Ni_7$ and the pseudo-binary $La_{1.5}Mg_{0.5}Ni_7$ [33]. They concluded that the presence of Mg improves the stabilization of the metal sublattice leading to the absence of amorphization upon solid-gas cycling.

The following point can also explain the better cycling stability of the Mg-substituted compounds. As mentioned in the introduction, the $R_A/R_B$ ratio determines the ability of $AB_2$ phases to absorb reversibly hydrogen [21, 16]. If this ratio is larger than 1.225 theoretically (or 1.37 experimentally), hydrogen induced amorphization occurs. As $R_{Sm}/R_{Ni}$ ≈ 1.45, we can expect a low cycling stability for $Sm_2Ni_7$. Since Mg is smaller than Sm, the $R_A/R_B$ ratio is reduced for the substituted $Sm_{1.65}Mg_{0.35}Ni_7$. This explains the enhanced cycling stability of the Mg-substituted compound. In addition, Mg substitutes Sm in the [$A_2B_4$] unit only, leading to a larger decrease of the $c$-axis compared to $a$-axis (Table 1). It is admitted that a smaller $c/a$ ratio induces less volume change during hydrogenation and consequently a better cycling stability [10].

3. **Effect of the nature of the phase ($A_2B_7$ or $A_5B_{19}$)**

**Cycling stability.** Upon hydrogenation, both $Sm_2Ni_7$ and $Sm_5Ni_{19}$ exhibit two pressure plateaus. The first pressure plateau is shorter and higher for $Sm_5Ni_{19}$ compound (Figure 2). The additional [$AB_5$] unit in the 5:19 explains the shorter plateau. For the Mg-free $AB_y$ compounds, the first hydrogenation step consists in the filling of [$A_2B_4$] units with H [41, 42, 43, 44, 45]. At the end of the first pressure plateau, the basic stacking period can be written as follow: $2x[AB_5]+[A_2B_4H_x]$ for $A_2B_7$ and $3x[AB_5]+[A_2B_4H_x]$ for $A_5B_{19}$. The amount of absorbed H is then: $x/18$ H/$M$ for $A_2B_7$ and $x/24$ H/$M$ for $A_5B_{19}$. Consequently, at the end of the first plateau, the H content is 4/3 larger for $A_2B_7$. This is in good agreement with our *P-c* isotherms (Figure 2): at the end of the first plateau 0.33 H/$M$ were absorbed by $Sm_5Ni_{19}$ and 0.41 H/$M$ were inserted in $Sm_2Ni_7$.

The fact that the pressure plateau is higher for 5:19 can be explained by its smaller average unit volume compared to 2:7 [37]. In addition, it might also be explained by the presence of one more [$AB_5$] unit in the basic stacking period of 5:19 compared to 2:7. As explained earlier, the first plateau corresponds to the absorption of H in the [$A_2B_4$] unit leading to an anisotropic expansion. The presence of one additional [$AB_5$] in the 5:19 prevent the expansion of the lattice. More energy is then needed to fill the [$A_2B_4$] unit, which would explain the higher plateau.

Another significant difference between the *P-c* isotherms of $Sm_2Ni_7$ and $Sm_5Ni_{19}$ is that the latter can desorb hydrogen after full hydrogenation, whereas $Sm_2Ni_7$ does not. The *P-c*-isotherm of $Sm_2Ni_7$ (Figure 2) shows two plateaus but no desorption at the isotherm condition. Therefore, hydrogen is not desorbed and remains in the cell. Furthermore, the diffraction peaks of $Sm_2Ni_7$ get broader, indicating some amorphization of the structure whereas the diffraction peaks of $Sm_5Ni_{19}$ are almost unchanged after dehydrogenation. The solid-gas cycling performed under low hydrogen pressure highlights a loss in capacity for $Sm_2Ni_7$ that can be attributed to crystallinity losses (Figure 4). Those two points suggest that the cycling stability is better for $Sm_5Ni_{19}$. Electrochemical cycling behavior was studied for $Gd_5Ni_{19}$ and $Gd_2Ni_7$ and similar results were observed [26]. The fact that $Sm_5Ni_{19}$ undergoes less amorphization than $Sm_2Ni_7$, promoting cycling properties, can be explained by the difference in structure.

V. **Conclusions**

$Sm_2Ni_7$ and $Sm_5Ni_{19}$ binary compounds and $Sm_{1.65}Mg_{0.35}Ni_7$ and $Sm_{4.5}Mg_{0.5}Ni_{19}$ pseudo-binary compounds were successful synthesized. Their structures were determined by X-ray diffraction, showing that they crystallize in two polymorphic structures (rhombohedral and hexagonal). Systematic study of their thermodynamic and corrosion properties allowed highlighting the following points:

- For short corrosion times, $A_5B_{19}$ compounds and Mg-substituted compounds show improved resistance to corrosion.
- $Sm_5Ni_{19}$ is more stable upon cycling than $Sm_2Ni_7$, this is attributed to the larger number of [$AB_5$] units for $Sm_5Ni_{19}$,
- Substitution of Sm by Mg increases the cycling stability, this is due to the decrease of the ratio $R_A/R_B$,
- Substitution of Sm by Mg leads to a larger but higher first pressure plateau, because of the smaller cell volume. A large plateau means that for a small variation of pressure a large capacity is obtained. It is then possible to adjust the pressure in the electrochemical window by substitution by larger (to decrease the plateau pressure) or smaller atoms (to increase the pressure of the plateau).